\documentclass[twocolumn,showpacs,preprintnumbers,amsmath,amssymb,aps]{revtex4-2}

\usepackage{epsfig}
\usepackage{dcolumn}
\usepackage{bm}
\usepackage[dvipsnames]{xcolor}
\usepackage[normalem]{ulem}
\usepackage{natbib,hyperref}

\usepackage{multirow}
\usepackage{adjustbox}

\begin{document}

\preprint{\today} 

\title{Measurement of the hyperfine coupling constants and absolute energies of the $12s \ ^2S_{1/2}$, $13s \ ^2S_{1/2}$, and $11d \ ^2D_{J}$ levels in atomic cesium}

\author{Jonah A. Quirk$^{1,2}$, Amy Damitz$^{1,2}$, Carol E. Tanner$^3$ and D. S. Elliott$^{1,2,4}$}

\affiliation{%
   $^1$Department of Physics and Astronomy, Purdue University, West Lafayette, Indiana 47907, USA\\
   $^2$Purdue Quantum Science and Engineering Institute, Purdue University, West Lafayette, Indiana 47907, USA\\
   $^3$Department of Physics, University of Notre Dame, Notre Dame, Indiana 46556, USA\\
    $^4$School of Electrical and Computer Engineering, Purdue University, West Lafayette, Indiana 47907, USA
   }

\date{\today}

\begin{abstract}
We report measurements of the absolute energies of the hyperfine components of the $12s \ ^2S_{1/2}$ and $13s \ ^2S_{1/2}$ levels of atomic cesium, $^{133}$Cs.  Using the frequency difference between these components, we determine the hyperfine coupling constants for these states, and report these values with a relative uncertainty of $\sim$0.06\%.  We also examine the hyperfine structure of the $11d \ ^2D_{J}$ ($J=3/2, 5/2$) states, and resolve the sign ambiguity of the hyperfine coupling constants from previous measurements of these states.  We also derive new, high precision values for the state energies of the $12s \ ^2S_{1/2}$, $13s \ ^2S_{1/2}$ and $11d \ ^2D_{J}$ states of cesium.
\end{abstract}

\maketitle 

\section{Introduction}\label{sec:introduction}
Accurate atomic structure calculations of atomic wave functions are critical for the quantitative interpretation of measurements of atomic parity violation (APV)~\cite{DzubaFS89, BlundellJS91, BlundellSJ92, Derevianko00, DzubaFG01, JohnsonBS01, KozlovPT01, DzubaFG02, FlambaumG05, PorsevBD09, PorsevBD10, DzubaBFR12, RobertsDF2013}. For example, in the sum-over-states approach for calculating the electric dipole transition moment $\mathcal{E}_{PNC}$ due to the weak force interaction between the nucleons and the electrons of an otherwise forbidden transition, precise values of electric dipole matrix elements and of the weak Hamiltonian can be used to relate the experimentally-determined value of $\mathcal{E}_{PNC}$ to the weak charge of the nucleus $Q_w$. The quality of these atomic structure calculations is judged by their ability to produce reliable values of measured (or measurable) quantities, such as energy eigenstates of the atom, transition moments (particularly for electric dipole transitions), etc.  To evaluate the quality of the matrix elements of the weak Hamiltonian $H_w$, one often examines the hyperfine coupling constants $A_{\rm hfs}$ of the atomic states involved.  Both the weak Hamiltonian and the hyperfine interaction are sensitive to the electronic wave-function in the vicinity of the nucleus. Therefore, accurate theoretical methods for calculating $A_{\rm hfs}$ are of high importance to APV studies.

In a recent report~\cite{GingesVF17} of an \emph{ab initio} calculation of the ground state hyperfine splitting (hfs) of cesium, Ginges, Volotka, and Fritzsche reported a calculated hfs of the ground state $\Delta \nu_{\rm hfs;6s} = 9177.4$ MHz, differing from the defined CODATA value of $9 \: 192.631 \: 770$ MHz by only 0.17\%. This relativistic Hartree-Foch many-body calculation includes effects of core polarization, correlation corrections, quantum electrodynamic (QED) radiative corrections (self-energy and vacuum polarization), and the effect of the nonuniform density of the magnetization of the nucleus, known as the Bohr-Weisskopf (BW) correction.  In $^{133}$Cs, the QED correction is –0.38\%, while the BW correction is –0.18\%, emphasizing the importance of these corrections towards the goal of achieving an uncertainty of 0.1–0.2\%. Ginges and Volotka~\cite{GingesV18} later proposed a method in which one uses the results of precise hfs measurements of excited $ns \ ^2S_{1/2}$ states to greatly improve the ground $6s \ ^2S_{1/2}$ state and $7s \ ^2S_{1/2}$ state hyperfine intervals. (From this point forward, we will use the abbreviated notation $ns$ in place of $ns \ ^2S_{1/2}$, $n p_J$ for $np \ ^2P_{J}$, and $n d_J$ for $nd \ ^2D_{J}$ states.)  They noted that the correlation corrections decreased with increasing principal quantum number $n$, approaching a constant but non-zero value.  They proposed to use measurements of the hfs in high ($n>9$) $ns$ states to determine the BW and QED corrections in these states, which can then be scaled for application to the $6s$ and $7s$ states. This removes the large uncertainties due to the BW and QED corrections from the hfs calculations. In a 2019 report, Grunefeld, Roberts, and Ginges~\cite{GrunefeldRG2019} examined trends in the corrections to the hyperfine coupling constants $A_{\rm hfs}$, to make predictions of these constants for $ns$ and $np_{1/2}$ states of cesium, where $6 \le n \le 17$, which they believe to be accurate at the 0.1\% level.

The hyperfine coupling constants $A_{\rm hfs}$ for $ns$ states of cesium for the lowest energy states (principle quantum numbers $6 \leq n \leq 17$) have been measured previously~\cite{GilbertWW83,FendelBUH07,StalnakerMGFDHT10,WuLWLC13,JinZXWMZXJ13,YangWYW16,TsekerisGHBS1974,TsekerisG1975,Tsekeris1976,FarleyTG1977}.  
In several of these works~\cite{FendelBUH07,StalnakerMGFDHT10,WuLWLC13,JinZXWMZXJ13} for low $n$ states, $6 \leq n \leq 9$, the researchers used a frequency comb source as a frequency reference, or, in some cases, even used the frequency comb source directly to excite the lines, and determine the absolute frequency of individual hyperfine lines.
The precision of these values of $A_{\rm hfs}$ are well below 0.1\%, and the  measurements by various groups are in good agreement with one another~\cite{GilbertWW83,FendelBUH07,StalnakerMGFDHT10,WuLWLC13,JinZXWMZXJ13,YangWYW16}. 
For states $n > 9$, the measurements~\cite{TsekerisGHBS1974,TsekerisG1975,Tsekeris1976,FarleyTG1977} date back to $\sim$1975, and the uncertainties are in the range 0.4-2.0\%. These measurements used a level-crossing technique, in which the investigators applied a static magnetic field as well as an r.f.~magnetic field, and detected a change in the fluorescence intensity or polarization at particular magnetic fields, which indicated that different energy states were Zeeman-tuned into resonance with the r.f.~transition.

In this paper, we report new, high-precision measurements of the hyperfine coupling constants for the $12s$ and $13s$ states of atomic cesium.  This measurement is part of our ongoing investigations toward an improved value of the weak charge of atomic cesium~\cite{antypasm12013,choipnc2016,TohDTJE19}.  
Our measurements provide higher precision values for $A_{\rm hfs}$ and the state energies $E_{cg}$ than were measured previously for the $12s$ and $13s$ states.

We also report measurements of the hyperfine coupling constants of the $11d_{3/2}$ and $11d_{5/2}$ states, whose excitation energy is in the same vicinity as that of the $12s$ and $13s$ states.  Our measurements are able to resolve the ambiguity of the sign of $A_{\rm hfs}$ for these excited states, and provide higher precision values for one of the $A_{\rm hfs}$ values and the state energies.  

The paper is organized as follows. We first discuss the measurement of the hyperfine structure of the $12s$ and $13s$ states.  We describe the measurement technique for these measurements, and analysis of the data, and compare our results with previous measurements and with theoretical results.  In Section \ref{sec:11d}, we discuss our measurements of $A_{\rm hfs}$ for the $11d_{3/2}$ and $11d_{5/2}$ states.  We follow this with a few concluding remarks.

\section{\texorpdfstring{$12s \ ^2S_{1/2} $ and $  13s \ ^2S_{1/2}$ measurements}{12s 2s1/2 and 13s 2s1/2 measurements}\label{sec:ns}}

\subsection{Experimental Configuration and Procedure}\label{sec:Measurement}
\begin{figure}
    \centering
    \includegraphics[width=0.35\textwidth]{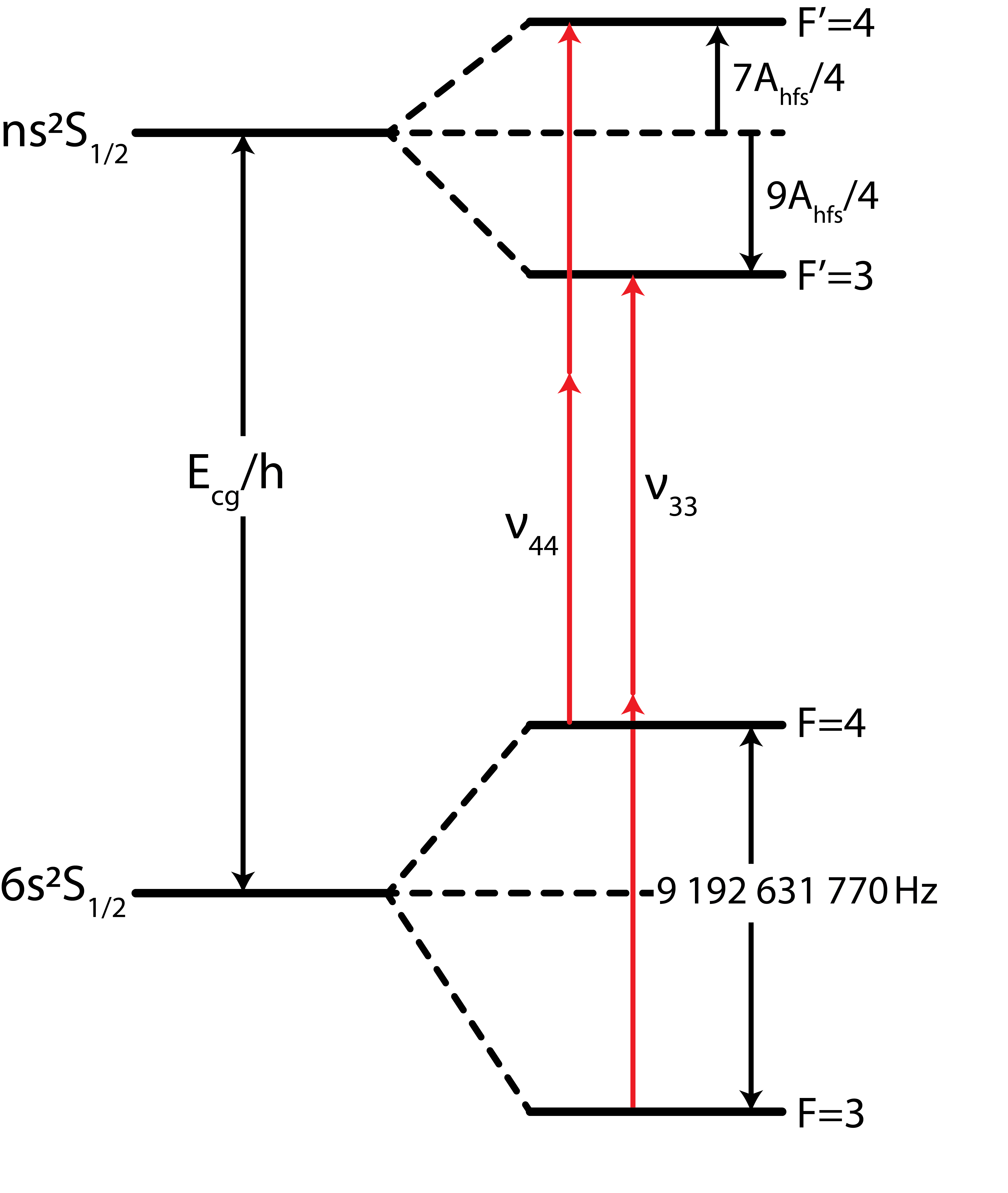}
    \caption{Energy level diagram showing the hyperfine components (not to scale) of the $6s$ and $ns$ states of cesium, where $n=12$ or $13$.  $\nu_{33}$ ($\nu_{44}$) indicates the frequency of the laser when resonant with the $F=3 \rightarrow F^{\prime}=3$ ($F=4 \rightarrow F^{\prime}=4$) two-photon transition.  $E_{\rm cg}$ is the energy of the $12s$ or $13s$ state in the absence of the hyperfine interaction (that is, the center-of-gravity of the state).}
    \label{fig:E_level_diagram}
\end{figure}

These measurements require precise determinations of the frequencies of the individual hyperfine components of the $6s \rightarrow 12s$ or $6s \rightarrow 13s$ transitions.  To achieve these, we use Doppler-free two-photon absorption in a cesium vapor, using a cw narrow-band laser, with precise calibration of the laser frequency using a frequency comb laser source.  Each transition consists of two well-resolved hyperfine lines, $F=3 \rightarrow F^{\prime}=3$ and $F=4 \rightarrow F^{\prime}=4$, where $F$ ($F^{\prime}$) is the total angular momentum of the $6s$ ground state ($12s$ or $13s$ excited state).  (Only $\Delta F = 0 $ transitions are allowed for this transition when the two photons are equal in frequency.)  We label the laser frequencies for these transitions $\nu_{33}$ and $\nu_{44}$, respectively, as shown in the energy level diagram of Fig.~\ref{fig:E_level_diagram}.  
These frequencies can be written
\begin{equation}\label{eq:nu33}
   \nu_{33} = \frac{1}{2} \left\{ \frac{E_{cg}}{h}  + \frac{9}{4} \left(  A_{\rm hfs, 6s} -  A_{\rm hfs, ns}\right) \right\}   
\end{equation}
and 
\begin{equation}\label{eq:nu44}
   \nu_{44} = \frac{1}{2} \left\{ \frac{E_{cg}}{h}  - \frac{7}{4} \left(  A_{\rm hfs, 6s} -  A_{\rm hfs, ns}\right) \right\}   ,
\end{equation}
where $E_{\rm cg}$ is the energy difference between the centers of gravity of the $6s $ and $ns$ states, and $h$ is the Planck constant. The factor $1/2$ in these expressions is included since we excite the transitions through two-photon absorption.  Through measurements of $\nu_{33}$ and $\nu_{44}$, and using the defined value for $A_{\rm hfs, 6s} = (\frac{1}{4})\times 9 \: 192.631 \: 770$ MHz, we are able to determine precise values for $A_{\rm hfs, ns}$ and $E_{cg}/h$. This measurement procedure is similar to that used in a previous work~\cite{WuLWLC13} for measurement of $A_{\rm hfs}$ of the $8s$ state, in which 10 kHz precision in the difference between hyperfine peaks was achieved.

 \begin{figure*}[t!]
\begin{centering}
\includegraphics[width=0.8\textwidth]{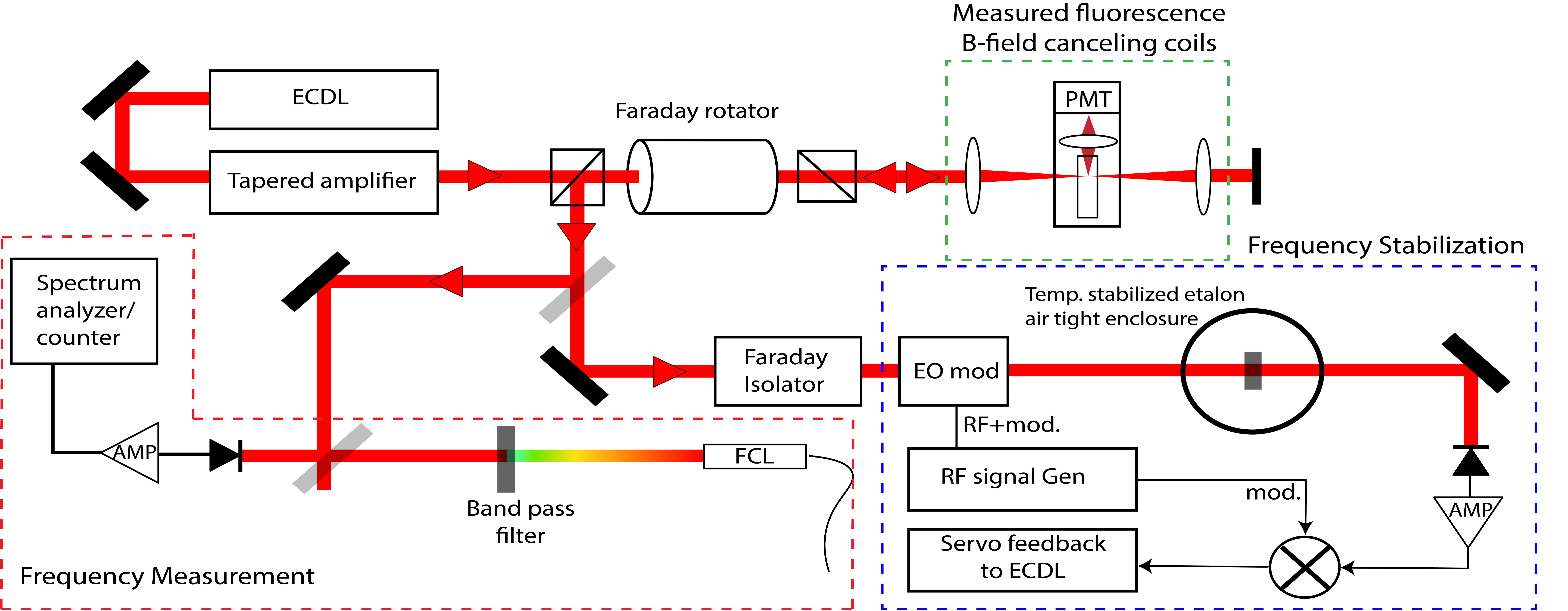}
 	  \caption{Experimental setup for the measurement of the two-photon absorption spectra.  The commercial diode laser (ECDL) and tapered amplifier generate 180-300 mW of narrow-band cw light, which is focused into a heated cesium vapor cell.  After passing through the cell, the laser light is reflected back on itself for Doppler-free two photon excitation.  We collect the fluorescence light (green box) emitted from the final $6p_{3/2} \rightarrow 6s$ step of the decay, which we measure with a photomultiplier tube (PMT).  We use a Faraday isolator to separate the retro-reflected beam from the input beam, while maintaining the linear polarization of the excitation beam in the vapor cell.  We stabilize the laser frequency (blue box), offset with an electro-optic modulator, to the transmission peak of a temperature-stabilized etalon.  We measure the frequency (red box) of the beat note between the laser light and a single tooth of a frequency-comb laser (FCL) for absolute calibration of the laser frequency.}
 	  \label{fig:sweepsetup}
\end{centering} 	  
\end{figure*}

We show a schematic layout of the experimental setup in Fig.~\ref{fig:sweepsetup}. A commercial external cavity diode laser (ECDL) and tapered amplifier in a master oscillator power amplifier (MOPA) configuration produce approximately 180-300 mW of cw narrow-band optical power near 670 nm.  The excitation wavelengths for the $12s$ and $13s$ states are 674.11 nm and 665.87 nm, respectively. We focus this light into a cesium vapor cell in a double-pass geometry to excite the cesium atoms through Doppler-free two-photon excitation. 
We use a Faraday isolator to separate the retro-reflected beam from the input beam, while preserving the linear polarization of the laser light in the vapor cell.  
The laser beam rejected by the Faraday isolator serves two purposes.  First, 
we use this beam to stabilize the laser frequency.  We achieve this by placing phase-modulation sidebands (40-110 MHz) on the beam using a broadband electro-optic modulator, dithering the sideband frequency (50 kHz dither frequency), generating an error signal from the transmission peak of one sideband through a 9.3 GHz FSR, temperature-stabilized etalon (mixed with 50 kHz and low pass filtered), and locking the laser frequency with this error signal. We tune the laser frequency indirectly by tuning the sideband frequency. Second, we use this beam to determine the laser  frequency $\nu$ throughout the duration of a laser scan.  We achieve this by combining the laser beam with the output of a frequency comb laser source on a beamsplitter, and measuring the absolute value of the beat frequency $\nu_{\rm beat} = \nu - \nu_{\rm FCL}$ between the laser (unmodulated) and a single tooth of the output of the frequency comb laser (of frequency $\nu_{\rm FCL}$) using a spectrum analyzer that is referenced to a GPS conditioned 10 MHz clock (Endrun Meridian).

The frequency comb laser is a commercial femtosecond erbium-doped fiber laser (Menlo FC1500), which when frequency doubled to 780 nm and spectrally broadened in a photonic crystal fiber (PCF), produces a coherent comb of light with a tooth spacing of $\nu_{\rm rep}=250$ MHz and an offset of $\nu_{\rm offset}=40$ MHz. Both the repetition rate $\nu_{\rm rep}$ and offset frequency $\nu_{\rm offset}$ are locked to the 10 MHz reference clock.
The absolute frequency of the laser is given by 
\begin{equation}\label{eq:laserfreq}
\nu = N \nu_{\rm rep} + \nu_{\rm offset} + \nu_{\rm beat},
\end{equation}
where the integer $N$ is the mode number, which labels the specific tooth of the frequency comb laser that we are beating against. We determine $N$ by measuring the laser frequency with a wavemeter whose accuracy is better than half the repetition rate, and determine the sign of the beat frequency $\nu_{\rm beat}$ by observing whether the beat frequency increases or decreases with increasing laser frequency.  

The vapor cell for these measurements is a fused silica cell, of dimensions $1 \times 1 \times 4.4$ cm$^3$, purchased for these measurements from Precision Glassblowing.  The cell fabricator used the following procedures to ensure high purity of the cell; purchased the highest-purity cesium, baked the cell at 425 $^\circ$C at $10^{-8}$ Torr for greater than 24 hours, repeatedly heated and transferred the alkali to the cell, and kept the cell under vacuum while sealing. During the course of a measurement, we maintain the temperature of the cell cold finger to within 0.1 $^{\circ}$C, with the cell windows at a higher temperature to avoid cesium condensation there.  We reduce the influence of collisions with the cell walls by collecting fluorescence from the central 6 mm region of the cell.  We cancel the local magnetic field at the cell location to a level below 10 mG in each direction with three pairs of current-carrying wire loops.  We also use these loops to intentionally apply a magnetic field of $\sim$1 G to the cell, with no observable effect on the spectra.  

We collect the fluorescence emitted by the excited state atoms with a one inch focal length, one inch diameter lens, positioned two inches from the interaction region.  
There are multiple decay routes that the atoms can follow as they relax to the ground state.  We choose to detect the 852 nm fluorescence from the $6p_{3/2} \rightarrow 6s$ decay, due to its large branching ratio ($>30$\%)~\cite{safronovadatabase}, the ability to discriminate the fluorescence from scattered laser light with an interference filter (peak transmission = 95\%, bandwidth = 10 nm), and the sensitivity of our available photomultiplier (PMT, spectral response R928) to this wavelength. We use an aperture at the image plane of the lens to further reduce the scattered light reaching the photomultiplier.  The 1 kV bias voltage applied to the PMT produces a PMT gain of $\sim1 \times 10^{7}$.  We observe the PMT output on an oscilloscope, and measure its value with a 16-bit National Instruments analog-to-digital converter (ADC). The time constant of the detection system as determined by the ADC input stage and an external capacitor is $\sim$250 ms.   
As we scan the laser frequency over the two-photon resonance, we record the fluorescence signal and the beat frequency at a rate of 1000 samples per second.

\subsection{Data Analysis}\label{sec:DataAnalysis}
\begin{figure}
    \centering
    \includegraphics[width=0.4\textwidth]{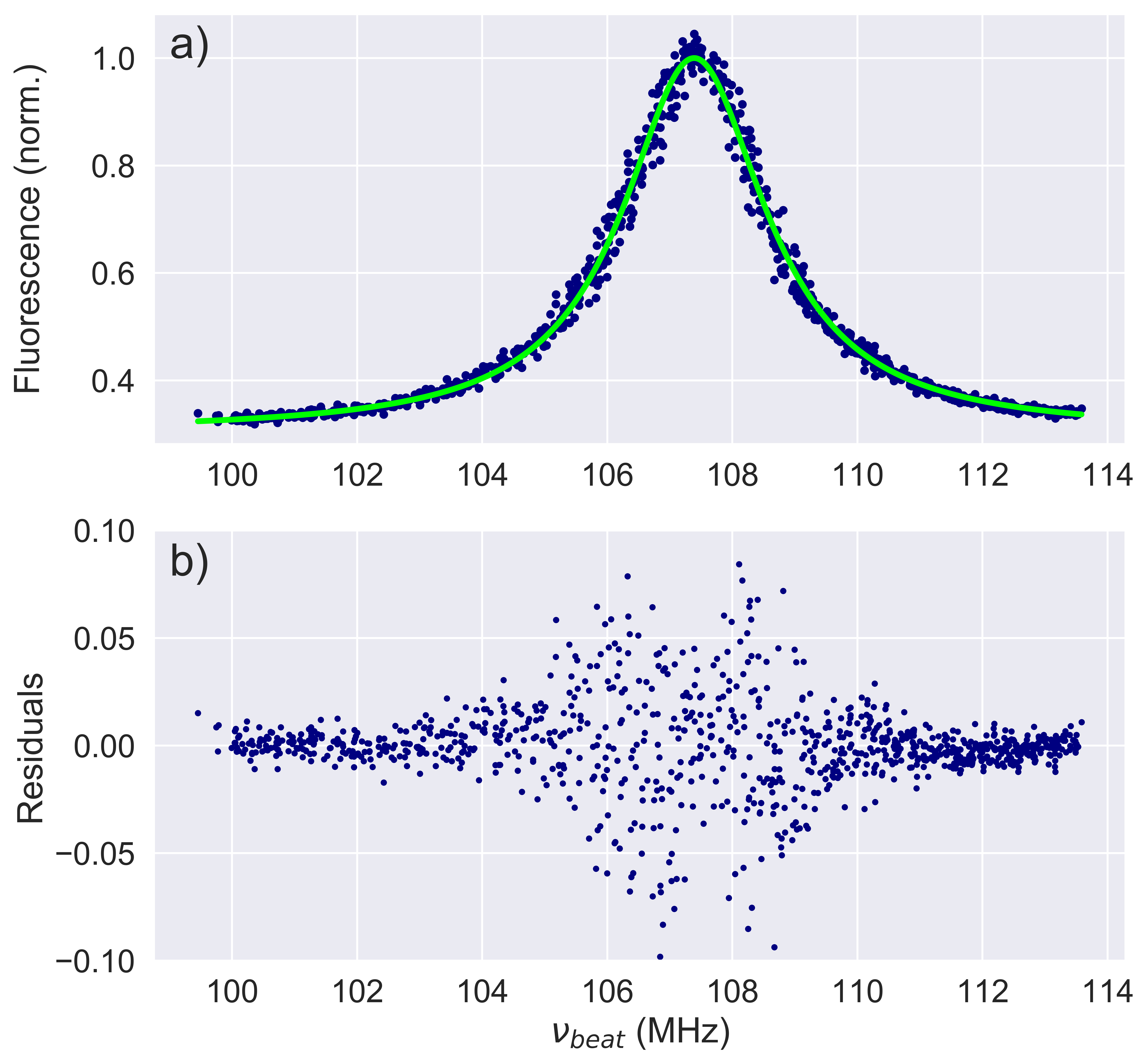}
    \caption{(a) An example of a two-photon spectrum of a single hyperfine line, consisting of the normalized fluorescence signal versus the beat frequency $\nu_{\rm beat}$.  Each data point is the signal collected in a 100 ms window as the laser frequency is scanned continuously over the 14 MHz span.  The solid green line is the result of a least-squares fit of a Lorentzian function to the data.  (b) The residuals show the difference between the data points and the fitted function. }\label{fig:fluorescence}
\end{figure}

We show a single spectrum of the $6s, F=4 \rightarrow 13s, F^{\prime} = 4$ line, as a representative example, in Fig.~\ref{fig:fluorescence} (a).
This spectrum shows the fluorescence signal, normalized to a peak value of 1 (actual peak voltage $\sim$100 mV), versus the laser beat frequency. This spectrum represents 400 seconds of data collection ($\sim$75\% of which is dead time to allow for data transfer), collected while scanning the laser frequency back and forth a total of four times. Each data point represents the signal averaged over 100 ms. The baseline in these data is primarily due to dark current ($<12$ mV dark current signal at high temperature, consistent with the specifications for the PMT) and scattered laser light, $<20$ mV signal.
For each spectrum, we perform a nonlinear, least-squares fit to a Lorentzian function utilizing the Levenberg-Marquardt algorithm included in the software package OriginPro.  Fitting parameters include: line center, linewidth, offset, and peak height.  We show the result for this fit for the data as the solid green line in Fig.~\ref{fig:fluorescence}(a).
A Lorentzian fit to the data provides, in each of our spectra, a good fit to the data.  

We show the residuals, {\emph i.e.} the difference between the data and the fitted function, in Fig.~\ref{fig:fluorescence}(b). At line center, the noise is $\sim$1\% of the peak level, consistent with Poissonian counting statistics.  On either side of line center, the noise level increases to $\sim$3-4\%, which we believe is due to frequency fluctuations of the laser field~\cite{AndersonJCSERZ1990}.  Power fluctuations of the laser beam are less than 0.1\% of the d.c.~power, and do not contribute significantly to these residuals.  

\begin{figure}
    \centering
   \includegraphics[width=0.45\textwidth]{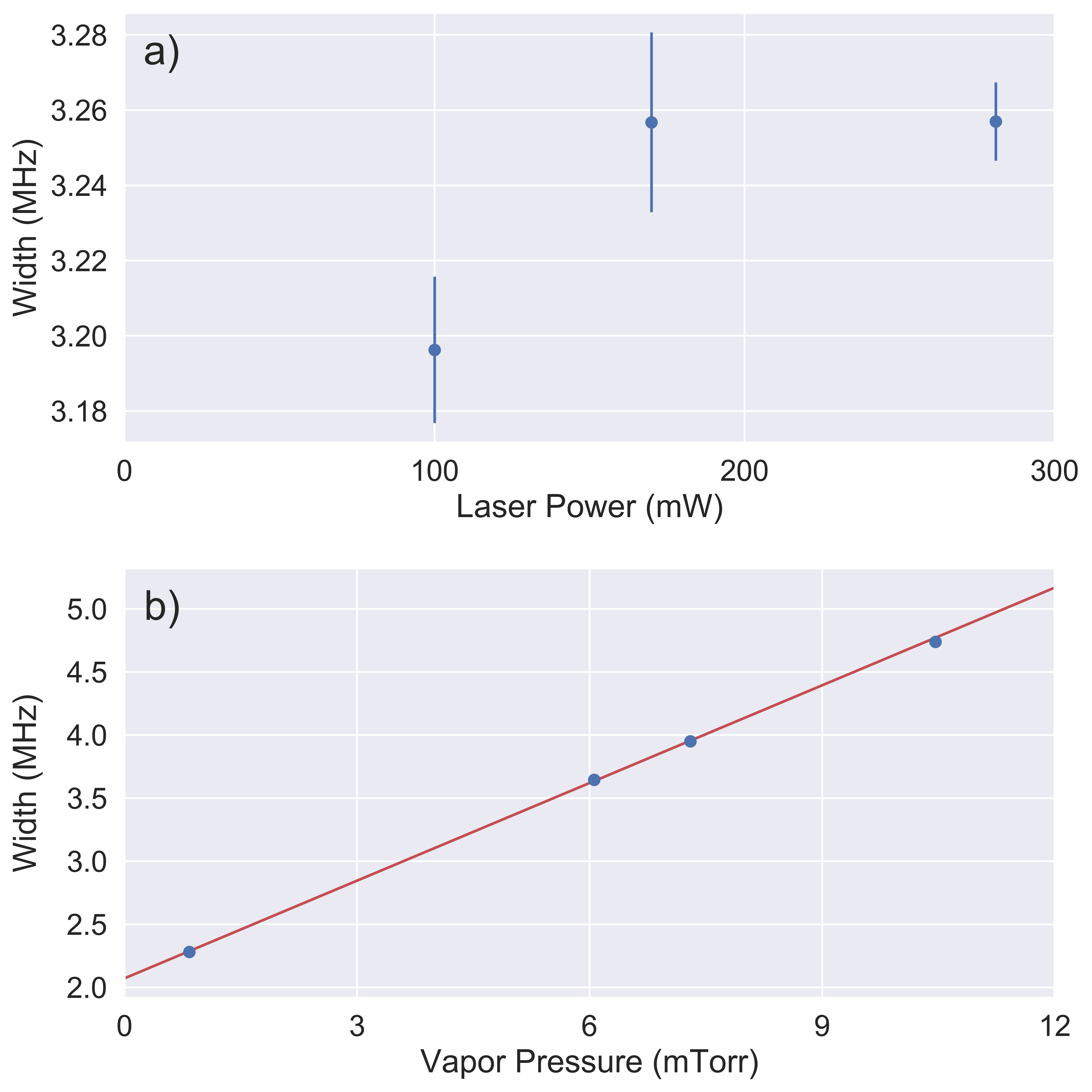}
    \caption{Plots of the linewidth of the two-photon $6s, F=3 \rightarrow 13s, F=3$  spectrum versus (a) the laser power, and (b) the cesium pressure.  }\label{fig:linewidth}
\end{figure}

The linewidth of the spectra, ranging from 2 - 4 MHz (see Fig.~\ref{fig:linewidth}), is the result of several factors, including contributions from the natural linewidth of the transition, collisional broadening, power broadening, transit time broadening, the 400 kHz linewidth of the laser, and residual Doppler broadening. The natural lifetime broadening is only a minor contributor, as the lifetimes of the 12s and 13s states are $\tau_{12s} = 573 \ (7)$ ns and $\tau_{13s} = 777 \ (8)$ ns~\cite{NeilA84}, corresponding to natural linewidths of 138 kHz and 103 kHz, respectively.  (We state these linewidths in terms of the laser frequency, which because of the two-photon excitation, is half the atomic frequency width.) The role of power broadening and collisions on the absorption linewidth can be seen clearly in Figs.~\ref{fig:linewidth}(a) and (b), respectively.
These plots show that collisional effects contribute to the linewidth, while power broadening is not significant.  We also expect transit time broadening to be significant for these measurements, since we used a moderately short focal length lens (7.5 cm) to focus the laser beam into the vapor cell in order to enhance the signal strength.  We estimate the beam radius at the focus to be $w \sim$8.0 $\mu$m, resulting in an estimated broadening of a few MHz.  Finally, we expect that some residual Doppler broadening, possibly resulting from imperfect alignment of the counter-propagating laser beams in the vapor cell, could result in some additional line broadening as well.  (This last effect is difficult to quantify, but its contribution to our signal seems possible, and we mention it for completeness.)

We repeat each measurement 6-9 times, and determine the mean and uncertainty in the line center from the distribution among these fits.  Typical values for the uncertainty (one standard error of the mean) in the line center range from 2-6 kHz.

\subsection{Results}\label{sec:Results}
\begin{figure}[b!]
    \centering
    \includegraphics[width=0.45\textwidth]{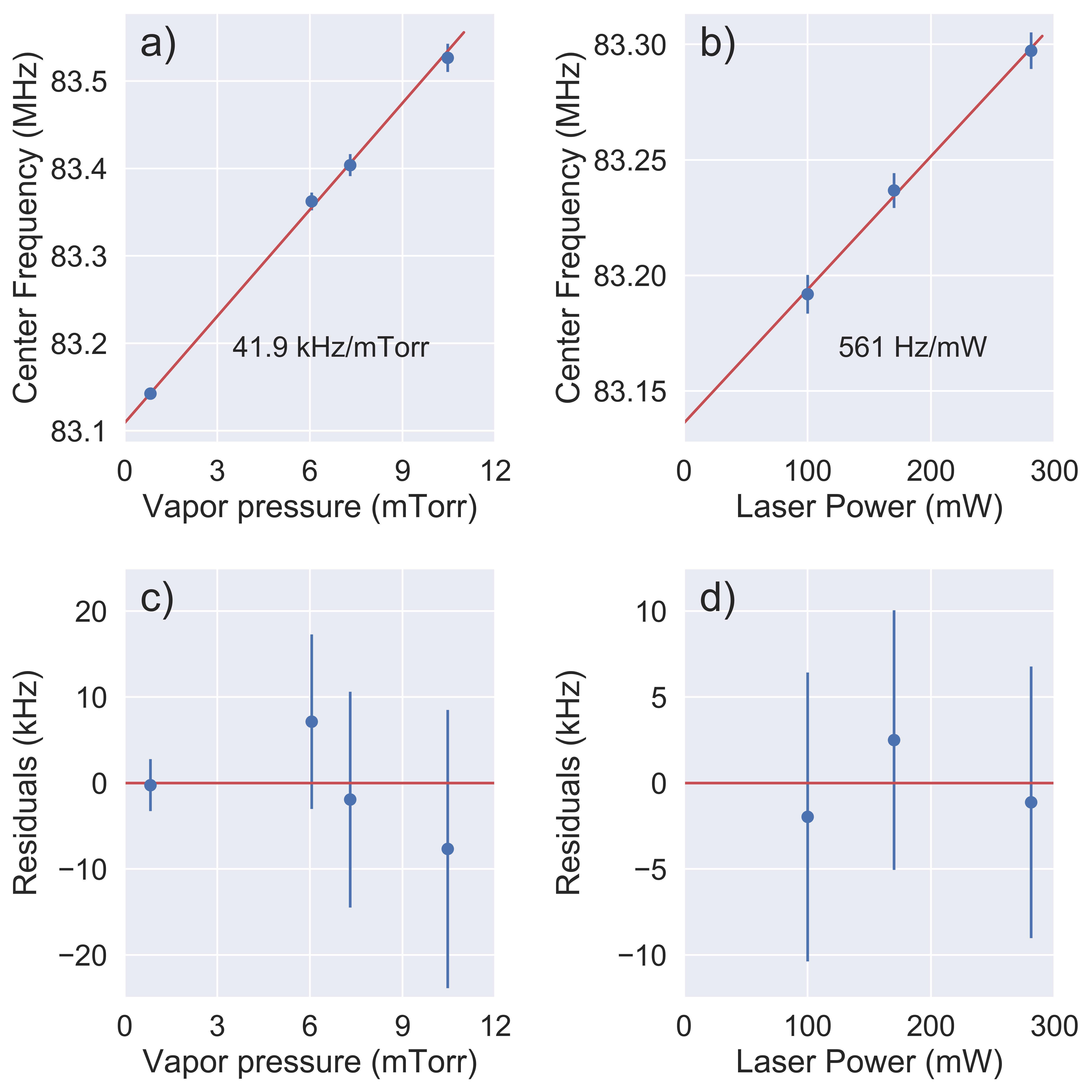}
    \caption{Plots of the line center of the two-photon $6s, F=3 \rightarrow 13s, F=3$ spectrum versus (a) the cesium density, and (b) the laser power.  We show the residuals between the data and a linear fit in plots (c) and (d).}\label{fig:linecenter}
\end{figure}

Collisions and power are expected to influence the line center of the spectra, so we measure the line center of each transition at a variety of vapor densities and laser powers.  In Table \ref{table:experiment_characteristics}, we list the ranges of amplitudes, cell temperatures, laser powers, and the resulting linewidths of the spectra.
\begin{table}[t!]
    \caption{Experimental parameters, including the ranges of amplitudes of the fluorescence signal detected by the PMT, cell temperatures, laser powers, and spectral widths of the spectra. 
    }
    \def\arraystretch{1.2}
    \begin{tabular}{|c |c|c|c|c| }
        \hline

    Transition &  Amplitude  & Temp. & Power  &\rule{0.05in}{0in} Linewidth \rule{0.05in}{0in} \\
       \rule{0.1in}{0in} &   (mV)  & ($^\circ $C) &  (mW) & (MHz) \\
		\hline \hline
      \rule{0.05in}{0in}  $6s \rightarrow 12s$	\rule{0.05in}{0in}	&\rule{0.05in}{0in} 15-150 \rule{0.05in}{0in} & \rule{0.05in}{0in}102-156 \rule{0.05in}{0in} & \rule{0.05in}{0in}90-186\rule{0.05in}{0in} & 2.7-4.8 \\ \hline
        $6s \rightarrow 13s$ 	& 12-110  & 110-157  & 70-285 & 2.4-5   \\ \hline
        $6s \rightarrow 11d$		& 5-10  & 77-104  & 300 & ~2.6          \\

    \hline
    \end{tabular}
    \label{table:experiment_characteristics}
\end{table}
We display plots of the line center of the $6s, F=3 \rightarrow 13s, F^{\prime}=3$ spectrum as a function of cesium vapor pressure and laser power in Fig.~\ref{fig:linecenter}. 
The error bars in these plots show the standard error of the mean in the transition frequencies $\sigma_{\nu}$ due only to the scatter among the independent measurements of the line center frequencies.  The total uncertainty $\sigma_{\nu}^{\rm total}$ of these data points must also include additional uncertainties due to the cell temperature $\sigma_T$ and the laser power $\sigma_P$. ($\sigma_T \sim 0.7^{\circ}$C is limited by the precision of our thermocouple reader, and is greater than the 0.1$^{\circ}$C temperature stability of the cell.)  The total uncertainty $\sigma_{\nu}^{total}$ in the line center of the spectrum is the quadrature sum of the statistical uncertainty $\sigma_{\nu}$, the product $\sigma_T \frac{d \nu}{dT}$, and the product $\sigma_P \frac{d \nu}{dP}$. The dependence of the line center on the cell temperature $\frac{d \nu}{dT}$ is significant, while its dependence on the laser power $\frac{d \nu}{dP}$ is rather weak.  We note that the a.c.~Stark shift of the line center varies with alignment of the counter propagating laser beams and varies linearly with laser power. To mitigate any errors this might cause, alignment was fixed for an entire determination of $\nu_{33}$ or $\nu_{44}$.
We fit the data as a linear function of pressure (derived from the cell temperature) and laser power, and extrapolate to zero pressure and zero laser power to determine the intercept; that is, the line centers $\nu_{33}$ and $\nu_{44}$ of the transition frequencies of the $F=3 \rightarrow F^{\prime}=3$, and $F=4 \rightarrow F^{\prime}=4$ transitions, respectively. The reduced $\chi^2$ for these fits is in the range 0.93-1.58. For those cases for which $\chi^2 > 1$, we increase the uncertainty of the line center by a factor $\sqrt{\chi^2}$.

We tabulate the sources of error and their magnitudes in Table~\ref{table:sourcesoferror}.
\begin{table}[b!]
    \caption{Sources of error and the uncertainty resulting from each, for the determinations of line centers for each of the spectra.   
    We add the errors in quadrature to obtain the total uncertainty. 
    }
    \def\arraystretch{1.2}
    \begin{tabular}{|l|  c| }
        \hline

      \rule{0.1in}{0in}Source             &   \rule{0.1in}{0in}$\sigma_{\rm int}$(kHz)	\rule{0.1in}{0in}	\rule{0in}{0.15in}  \\
		\hline \hline
       \rule{0.1in}{0in} Fit, $\sigma_{\nu}^{\rm total}$           	& 13-22   \\
		
       \rule{0.1in}{0in} FCL frequency, $\nu_{\rm FLC}$	& $<0.5$   \\
       \rule{0.1in}{0in} Zeeman			            	& $<0.3$ 	 \\

                                        \hline \hline
        \rule{0.1in}{0in}\rule{0in}{0.15in}Total uncertainty, $\sigma_{\rm int}^{\rm total}$	\rule{0.1in}{0in}			& 13-22\\
    \hline
    \end{tabular}
    \label{table:sourcesoferror}
\end{table}
The primary contribution comes from the statistical determination of the line center $\sigma_{\nu}^{\rm total}$, and is listed as `Fit.'  We derive this uncertainty for each peak of the spectrum, using the data at the various laser powers and cell temperatures, and extrapolating to zero laser power and zero cell density.  $\nu_{\rm FLC}$ is our estimate of fluctuations of the frequency of the FCL laser, based on the fractional stability of the GPS 10 MHz clock and the comb tooth number $N$. The Zeeman error is our estimate of maximum possible line shifts due to less-than-perfect cancellation of the magnetic field, and any resulting Zeeman shift, at the location of the vapor cell.

We present the results for the laser frequencies $\nu_{33}$ and $\nu_{44}$ in Table \ref{table:linecenterfreqs}.
\begin{table*}[t!]
\begin{center}
 \def\arraystretch{1.2}
 \begin{tabular}{|c|c|c|c|c|} \hline
\rule{0in}{0.15in}   &  &  & \multicolumn{2}{c|}{\rule{0in}{0.2in}$E_{cg}/h$ (MHz) } \\ \cline{4-5} 
Line & $\nu_{33}$ (MHz) &  $\nu_{44}$ (MHz)   & this work & Prior exp.  \cite{WeberS1987}   \\  \hline
\rule{0in}{0.2in} $6s \rightarrow 12s$  & \rule{0.1in}{0in} $ 444 \: 726 \: 731.369 \ (22)$ \rule{0.1in}{0in}  & \rule{0.1in}{0in} $444 \: 722 \: 187.689 \ (19) $ \rule{0.1in}{0in}  &$ 889 \: 448 \: 351.098 \ (29) $ &  $889 \: 448 \: 348.5 \ (60)$
  \\
 
\rule{0.05in}{0in}$6s \rightarrow 13s$    & $450 \: 227 \: 707.055 \ (13)$ &  $450 \: 223 \: 147.601 \ (15) $   & $900 \: 450 \: 284.724 \ (20)  $& $900 \: 450 \: 282.0 \ (60) $\\

\rule{0.1in}{0in}$6s \rightarrow 11d_{3/2}$\rule{0.1in}{0in}    & - &  -    &\rule{0.1in}{0in} $896 \: 269 \: 630.698 \ (65)$ \rule{0.1in}{0in} &\rule{0.1in}{0in} $896 \: 269 \: 624.7 \ (60) $ \rule{0.1in}{0in}
 \\

$6s \rightarrow 11d_{5/2}$    & - &  -    & $896 \: 365 \: 856.56 \ (24)$  & $896 \: 365 \: 852.6 \ (60)$

\\ \hline \hline 
\end{tabular}
\end{center} 
  \caption{Summary of results for the line centers of the hyperfine components of the $6s \rightarrow 12s$ and $6s \rightarrow 13s$ transitions, and the state energies $E_{\rm cg}/h$ of the $12s$, $13s$, $11d_{3/2}$, and $11d_{5/2}$ states of $^{133}$Cs.  The numbers in parentheses following each value are the $1 \sigma$ standard error of the mean in the least significant digits.}  \label{table:linecenterfreqs}
\end{table*}
Using Eqs.~(\ref{eq:nu33}) and (\ref{eq:nu44}), the weighted average of these two transition frequencies yields the energy of the center of gravity of the state
\begin{equation}
    E_{cg} = 2 \times \frac{h}{16}\left\{ 7 \nu_{33} + 9 \nu_{44} \right\}
\end{equation}
while the difference gives the hyperfine coupling constant
\begin{equation}
    A_{\rm hfs, ns} = A_{\rm hfs, 6s} - \frac{1}{2}\left\{  \nu_{33} - \nu_{44} \right\}.
\end{equation}
We have included $E_{cg}/h$ in Table \ref{table:linecenterfreqs}.
The values of the state energies $E_{cg}/h$ are in agreement with, but more precise by a factor of a few hundred than, the previous determination~\cite{WeberS1987}.  

We show our results for the hyperfine coupling constants $A_{\rm hfs, 12s}$ and $A_{\rm hfs, 13s}$ in Table \ref{table:hfsconstants}. These values are in agreement with the values measured using level-crossing spectroscopy~\cite{TsekerisG1975,Tsekeris1976}.  Our uncertainties are smaller by a factor of almost ten. There are two theoretical values of $A_{\rm hfs, 12s}$ available for comparison.  The authors of Ref.~\cite{TangLS2019} used the Dirac-Fock wavefunctions, with third-order many-body perturbation theory, and coupled-cluster method in single and double approximations.  Their result differs by 0.14\% from our value.  The theoretical calculations of Ginges, \emph{et al.}~\cite{GrunefeldRG2019} is in better agreement with our value, differing by  $<0.07\%$.  For $A_{\rm hfs, 13s}$, the result of Ref.~\cite{GrunefeldRG2019} is in similar good agreement with our value, differing by only 0.06\%.

\begin{table}[t!]
\begin{center}
 \begin{tabular}{|c|c|c|c|c|} \hline
    &  \multicolumn{4}{c|}{\rule{0in}{0.2in}$A_{\rm hfs} $ (MHz) } \\ \cline{2-5}
 State &   \multicolumn{2}{c|}{ Experiment\rule{0in}{0.15in}}   & \multicolumn{2}{c|}{ Theory} \\ \cline{2-5}   
  & This work & Prior exp. &\rule{0in}{0.15in}  Ref.~\cite{TangLS2019}  &  Ref.~\cite{GrunefeldRG2019} \\ \hline \hline
$12s$\rule{0in}{0.15in}    &  26.318 (15) &  26.31 (10) \cite{TsekerisG1975} &  26.28 &  26.30 (2)  \\ \hline
$13s$\rule{0in}{0.15in}    &  18.431 (10) & 18.40 (11) \cite{Tsekeris1976} & -- & 18.42 (1) \\ \hline

\multirow{2}{*}{$11d_{3/2}$} & \multirow{2}{*}{+1.0530 (69)} &  $\rule{0in}{0.15in} \pm$1.055 (15) \cite{SvanbergB1974}    & \multirow{2}{*}{ 1.06} & \multirow{2}{*}{--}  \\
  &  &   $\pm$1.04 (5) \cite{DeechLPS1977} & & \\ \hline
 $11d_{5/2}$\rule{0in}{0.15in}   & $-$0.21 (6) & $\pm$0.24 (6) \cite{SvanbergB1974} &  $-0.142$  & -- \\ \hline \hline 
\end{tabular}
\end{center} 
  \caption{Summary of results for the hyperfine coupling constants $A_{\rm hfs} $ of the $12s$, $13s$, $11d_{3/2}$ and $11d_{5/2}$ states of $^{133}$Cs. The numbers in parentheses following each value are the $1 \sigma$ standard error of the mean in the least significant digits.}    \label{table:hfsconstants}
\end{table}

\section{$11d \ ^2D_{3/2}$ and $11d \ ^2D_{5/2}$ measurements}\label{sec:11d}
\begin{figure}[b!]
    \centering
    \includegraphics[width=0.45\textwidth]{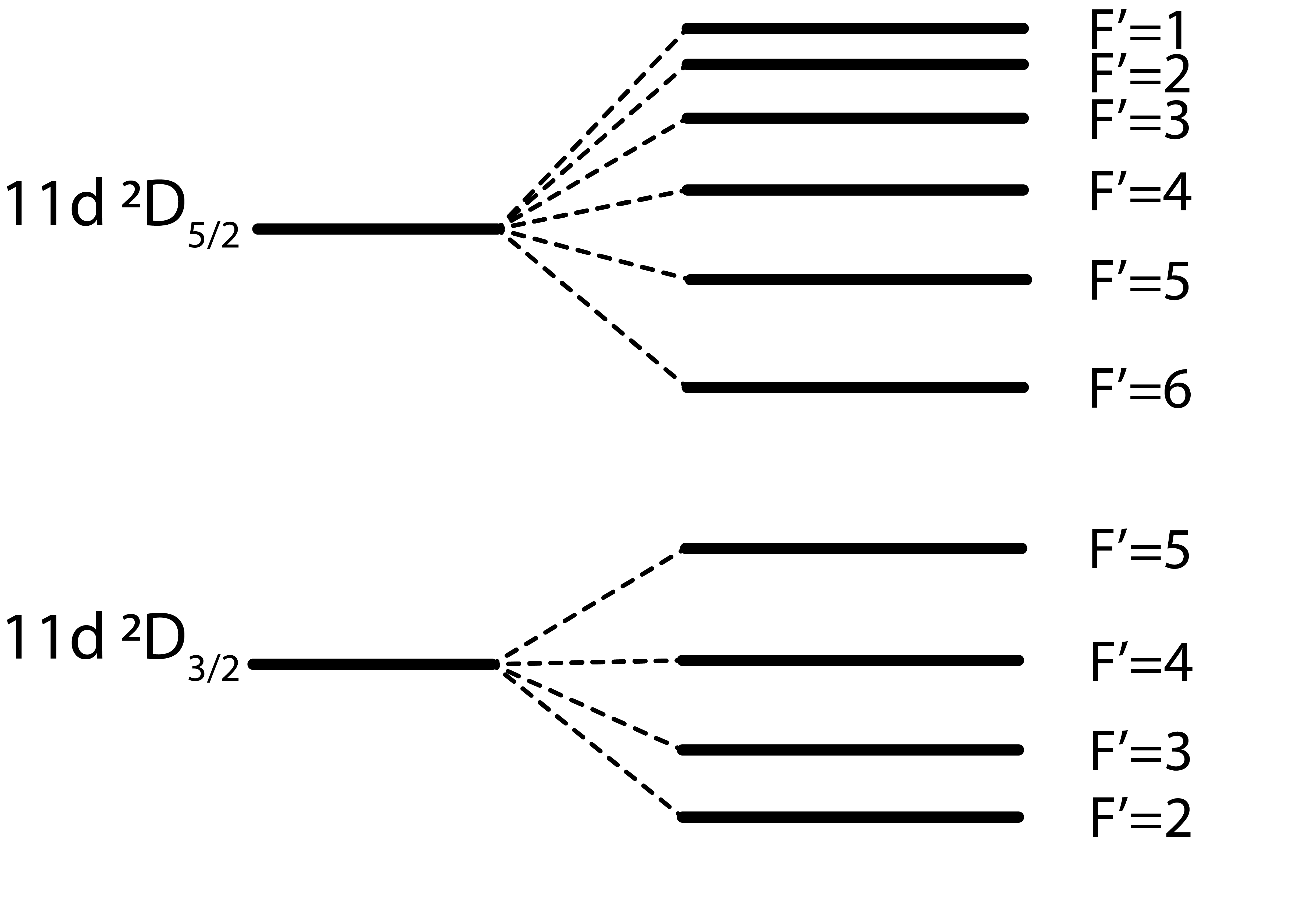}
    \caption{Energy level diagram showing the hyperfine components of the $11d_{3/2}$ and $11d_{5/2}$ states in cesium. Not shown here is the ground state from which we excite the cesium atoms. Note that the $11d_{5/2}$ state is inverted, with the level energy decreasing with increasing $F^{\prime}$. }\label{fig:e_level_diagram_11d}
\end{figure}

We used a similar procedure to measure the hyperfine structure of the $11d_{3/2}$ and $11d_{5/2}$ levels at a wavelength of $\lambda =$ 668.98 nm and 668.91 nm, respectively.  The most significant differences between these measurements and those of the 12s and 13s states are that the 11d lines are somewhat stronger, the hyperfine structure is more interesting (four or five hyperfine components within each spectrum), and the hyperfine splitting is much smaller.  We show an energy level diagram of the $11d_{3/2}$ and $11d_{5/2}$ states in Fig.~\ref{fig:e_level_diagram_11d}.

\begin{figure}
    \centering
    \includegraphics[width=0.45\textwidth]{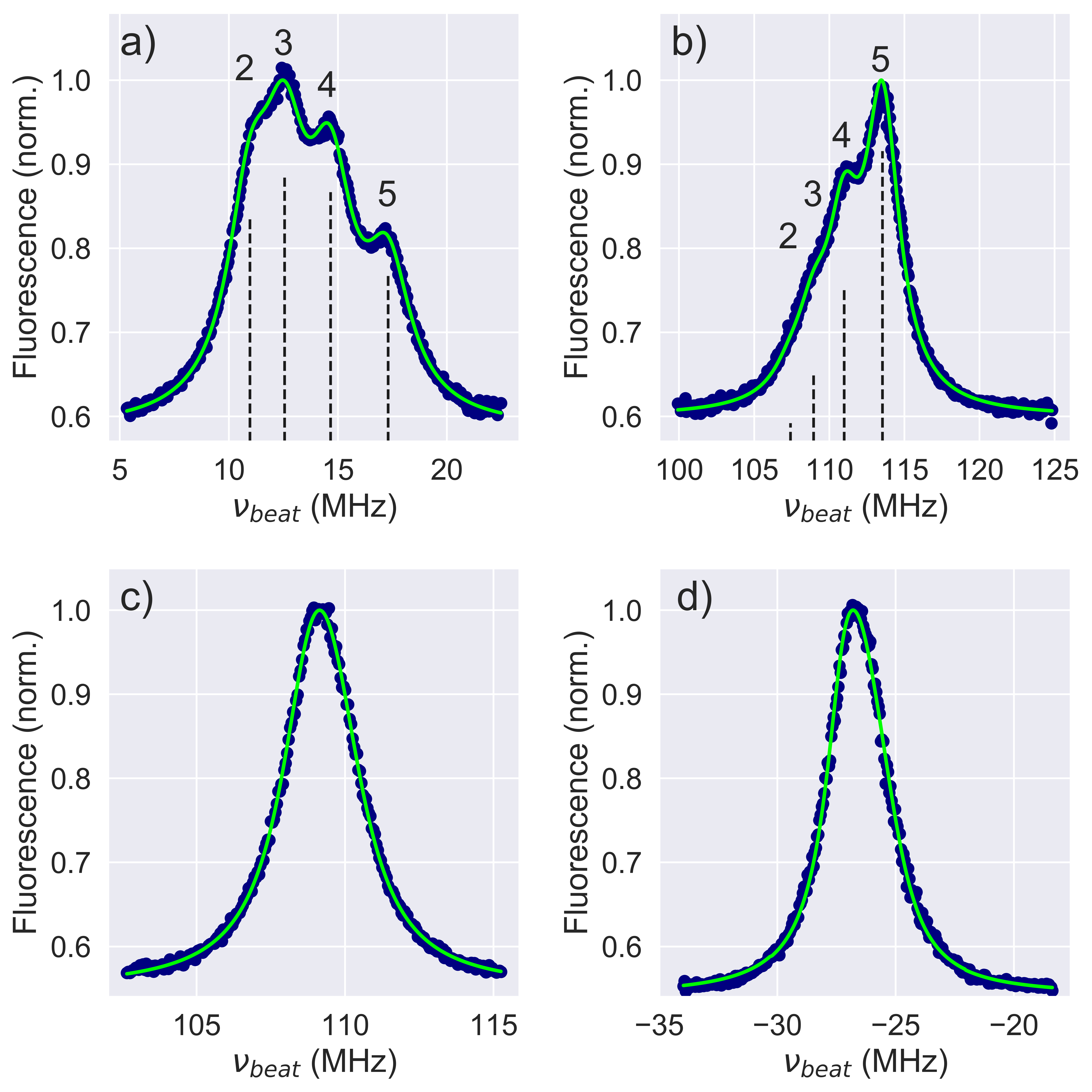}
    \caption{Spectra of the $11d$ states.  (a) $6s, F=4 \rightarrow 11d_{3/2}, F^{\prime}$, (b) $6s, F=3 \rightarrow 11d_{3/2}, F^{\prime}$, (c) $6s, F=4 \rightarrow 11d_{5/2}, F^{\prime}$, (d) $6s, F=3 \rightarrow 11d_{5/2}, F^{\prime}$,  The green curve is the result of a least squares fit to the spectra.  The vertical lines indicate the positions and relative linestrengths of each of the individual hyperfine components to the spectra.} \label{fig:11d}
\end{figure}

We show sample spectra in Fig.~\ref{fig:11d}. 
The upper two spectra are $6s, F \rightarrow 11d_{3/2}, F^{\prime}$, with $F$ = (a) 4 and (b) 3, while the lower spectra are $6s, F \rightarrow 11d_{5/2}, F^{\prime}$, with $F$ = (c) 4 and (d) 3.  The vertical lines in Figs.~\ref{fig:11d}(a) and (b) show the positions of the individual components of these transitions, with the height of the lines indicating the calculated relative strength of the transition, and $F^{\prime}$ labeled for each.  Some of the individual peaks in the $11d_{3/2}$ spectra are resolved. We fit a multi-component lineshape to the measured spectra, using computed values for the relative spacing and heights of the individual components, and show the results as the green solid lines in the figures.  The only adjustable parameters for these fits are the line center frequency, the linewidth of individual lines, the hyperfine coupling constant $A_{\rm hfs}$, the baseline, and an overall peak height.

For the $6s, F \rightarrow 11d_{5/2}, F^{\prime}$ spectra, shown in Fig.~\ref{fig:11d}(c) and (d), the spacing between the hyperfine components of the transition is much smaller than the linewidth of the individual peaks.  
There is, however, a slight asymmetry to the peaks, which we exploit to extract a negative value of $A_{\rm hfs}$ for the $11d_{5/2}$ state.
The peak asymmetry of the $6s, F=3 \rightarrow 11d_{5/2}$ lines and that of the $6s, F=4 \rightarrow 11d_{5/2}$ lines are reversed from one another, as expected based on the hyperfine line positions and calculated relative line strengths.  While we expect the latter spectrum to produce a more reliable value of $A_{\rm hfs}$ due to its larger asymmetry, we find consistent results for the two lines.  The result of the least-squares fits to the measured data are the solid green lines in Fig.~\ref{fig:11d}(c) and (d).  For these fits, we allowed the line center frequency, the hyperfine coupling constant $A_{\rm hfs}$, the baseline, and an overall peak height to adjust, but fixed the width of each individual lineshape at 2.6 MHz, which is the fitted value of the linewidth for the $12s$, $13s$, and $11d_{3/2}$ spectra under similar conditions of temperature and laser power.

We carried out these measurements at a much lower vapor density than we used for the $12s$ and $13s$ studies, in order to resolve the individual hyperfine levels in the $11d_{3/2}$ states. Working at this lower density while still maintaining measurable signal was allowed by the stronger $11d$ two-photon transition strength.  See Table \ref{table:experiment_characteristics}.  We measured the $A_{\rm hfs, 11d_{3/2}}$ coefficient 8-10 times for both high (cold finger temperature = $104 ^{\circ}$C) and low ($ 88 ^{\circ}$C) vapor pressure, where the vapor pressure varied by a factor of 3. We then extrapolated back to zero vapor pressure. The values of $A_{\rm hfs, 11d_{3/2}}$ at these two densities varied little, and the zero vapor pressure determination lies within the statistical spreads of the high and low vapor pressure fitted values of $A_{\rm hfs, 11d_{3/2}}$. We did not limit the laser power, as power broadening or power shifts were not evident in our measurements. We measured the spectra of the $6s \rightarrow 11d_{5/2}$ lines at only a single cell temperature (77$^{\circ}$C) to minimize the linewidth of the transition.

We present the results for the $A_{\rm hfs, 11d}$ in Table \ref{table:hfsconstants}.  Our uncertainty for $A_{\rm hfs, 11d_{3/2}}$ is 0.7\%, while for $A_{\rm hfs, 11d_{5/2}}$, which is smaller in magnitude, it is 27\%.  It is interesting to note that $A_{\rm hfs, 11d_{3/2}}$ is positive, while $A_{\rm hfs, 11d_{5/2}}$ is negative, in agreement with the theoretical values of Ref.~\cite{TangLS2019}.  Our results for $A_{\rm hfs, 11d}$ are in agreement with those of the previous measurement based on the level crossing technique by Svanberg and Belin~\cite{SvanbergB1974}.  The uncertainty of our measurement of $A_{\rm hfs, 11d_{5/2}}$ is the same as that of Ref.~\cite{SvanbergB1974}, while for $A_{\rm hfs, 11d_{3/2}}$, our uncertainty is smaller by more than a factor of two.  Agreement with the magnitude of the theoretical value for $A_{\rm hfs, 11d_{3/2}}$ of Ref.~\cite{TangLS2019} is good, differing by only 0.06\%. For $A_{\rm hfs, 11d_{5/2}}$, however, their value is a little more than one $\sigma$ smaller than our measured value.

In Table \ref{table:linecenterfreqs}, we report center-of-gravity energies for the $11d_{3/2}$ and $11d_{5/2}$ states $E_{cg}/h$, as determined from the fits described above.  The $6s \rightarrow 11d_{3/2}$ transition peaks are better resolved than the $6s \rightarrow 11d_{5/2}$ peaks (See Fig.~\ref{fig:11d}), and we were able to measure the temperature dependence and extrapolate back to zero vapor pressure.  The improved resolution and a greater number of data sets both contribute to a lower uncertainty in this line compared to $6s \rightarrow 11d_{5/2}$. The shift in line center frequency due to vapor pressure was  stronger in the $6s \rightarrow 11d_{3/2}$ transition than the $6s \rightarrow 12s$ and $6s \rightarrow 13s$ transitions, 158 kHz/mTorr instead of 36-42 kHz/mTorr. We assumed the shift in the $6s \rightarrow 11d_{5/2}$ transition was the same and used that value to estimate the zero pressure line center frequency, this shift was 16 kHz and was much less than the uncertainty in the line center. Our values agree well with the previous measurements \cite{WeberS1987}, but have $>25$ times lower uncertainty.

\section{Conclusion}\label{sec:Conclusions}
In this work, we have reported new, high precision measurements of the hyperfine coupling constants $A_{\rm hfs}$ of the $12s \ ^2S_{1/2}$ and the $13s \ ^2S_{1/2}$ states of atomic cesium.  In combination with previous measurements of $A_{\rm hfs}$ for lower $n$ states ($6 \le n \le 9$), these data add to development of atomic theory expected to be precise at the $\sim0.1$\% level.  We have also reported our measurements of the $A_{\rm hfs}$ for the $11d \ ^2D_{3/2}$ and $11d \ ^2D_{5/2}$ states.  Our measurements are in agreement with previous measurements, but resolve the ambiguity of the sign of $A_{\rm hfs}$. We also have reported new, higher precision values for the state energies of the $12s \ ^2S_{1/2}$, $13s \ ^2S_{1/2}$, and $11d \ ^2D_{J}$ states of cesium.

This material is based upon work supported by the National Science Foundation under Grant Number PHY-1912519.  We acknowledge useful conversations with J.~Ginges, and frequency comb laser advice from D.~Leaird and N.~O'Malley.

\bibliography{biblio}

 \newcommand{\noop}[1]{}
\begin{thebibliography}{36}%
\makeatletter
\providecommand \@ifxundefined [1]{%
 \@ifx{#1\undefined}
}%
\providecommand \@ifnum [1]{%
 \ifnum #1\expandafter \@firstoftwo
 \else \expandafter \@secondoftwo
 \fi
}%
\providecommand \@ifx [1]{%
 \ifx #1\expandafter \@firstoftwo
 \else \expandafter \@secondoftwo
 \fi
}%
\providecommand \natexlab [1]{#1}%
\providecommand \enquote  [1]{``#1''}%
\providecommand \bibnamefont  [1]{#1}%
\providecommand \bibfnamefont [1]{#1}%
\providecommand \citenamefont [1]{#1}%
\providecommand \href@noop [0]{\@secondoftwo}%
\providecommand \href [0]{\begingroup \@sanitize@url \@href}%
\providecommand \@href[1]{\@@startlink{#1}\@@href}%
\providecommand \@@href[1]{\endgroup#1\@@endlink}%
\providecommand \@sanitize@url [0]{\catcode `\\12\catcode `\$12\catcode
  `\&12\catcode `\#12\catcode `\^12\catcode `\_12\catcode `\%12\relax}%
\providecommand \@@startlink[1]{}%
\providecommand \@@endlink[0]{}%
\providecommand \url  [0]{\begingroup\@sanitize@url \@url }%
\providecommand \@url [1]{\endgroup\@href {#1}{\urlprefix }}%
\providecommand \urlprefix  [0]{URL }%
\providecommand \Eprint [0]{\href }%
\providecommand \doibase [0]{https://doi.org/}%
\providecommand \selectlanguage [0]{\@gobble}%
\providecommand \bibinfo  [0]{\@secondoftwo}%
\providecommand \bibfield  [0]{\@secondoftwo}%
\providecommand \translation [1]{[#1]}%
\providecommand \BibitemOpen [0]{}%
\providecommand \bibitemStop [0]{}%
\providecommand \bibitemNoStop [0]{.\EOS\space}%
\providecommand \EOS [0]{\spacefactor3000\relax}%
\providecommand \BibitemShut  [1]{\csname bibitem#1\endcsname}%
\let\auto@bib@innerbib\@empty
\bibitem [{\citenamefont {Dzuba}\ \emph {et~al.}(1989)\citenamefont {Dzuba},
  \citenamefont {Flambaum},\ and\ \citenamefont {Sushkov}}]{DzubaFS89}%
  \BibitemOpen
  \bibfield  {author} {\bibinfo {author} {\bibfnamefont {V.}~\bibnamefont
  {Dzuba}}, \bibinfo {author} {\bibfnamefont {V.}~\bibnamefont {Flambaum}},\
  and\ \bibinfo {author} {\bibfnamefont {O.}~\bibnamefont {Sushkov}},\ }\href
  {https://doi.org/https://doi.org/10.1016/0375-9601(89)90777-9} {\bibfield
  {journal} {\bibinfo  {journal} {Physics Letters A}\ }\textbf {\bibinfo
  {volume} {141}},\ \bibinfo {pages} {147 } (\bibinfo {year}
  {1989})}\BibitemShut {NoStop}%
\bibitem [{\citenamefont {Blundell}\ \emph {et~al.}(1991)\citenamefont
  {Blundell}, \citenamefont {Johnson},\ and\ \citenamefont
  {Sapirstein}}]{BlundellJS91}%
  \BibitemOpen
  \bibfield  {author} {\bibinfo {author} {\bibfnamefont {S.~A.}\ \bibnamefont
  {Blundell}}, \bibinfo {author} {\bibfnamefont {W.~R.}\ \bibnamefont
  {Johnson}},\ and\ \bibinfo {author} {\bibfnamefont {J.}~\bibnamefont
  {Sapirstein}},\ }\href {https://doi.org/10.1103/PhysRevA.43.3407} {\bibfield
  {journal} {\bibinfo  {journal} {Phys. Rev. A}\ }\textbf {\bibinfo {volume}
  {43}},\ \bibinfo {pages} {3407} (\bibinfo {year} {1991})}\BibitemShut
  {NoStop}%
\bibitem [{\citenamefont {Blundell}\ \emph {et~al.}(1992)\citenamefont
  {Blundell}, \citenamefont {Sapirstein},\ and\ \citenamefont
  {Johnson}}]{BlundellSJ92}%
  \BibitemOpen
  \bibfield  {author} {\bibinfo {author} {\bibfnamefont {S.~A.}\ \bibnamefont
  {Blundell}}, \bibinfo {author} {\bibfnamefont {J.}~\bibnamefont
  {Sapirstein}},\ and\ \bibinfo {author} {\bibfnamefont {W.~R.}\ \bibnamefont
  {Johnson}},\ }\href {https://doi.org/10.1103/PhysRevD.45.1602} {\bibfield
  {journal} {\bibinfo  {journal} {Phys. Rev. D}\ }\textbf {\bibinfo {volume}
  {45}},\ \bibinfo {pages} {1602} (\bibinfo {year} {1992})}\BibitemShut
  {NoStop}%
\bibitem [{\citenamefont {Derevianko}(2000)}]{Derevianko00}%
  \BibitemOpen
  \bibfield  {author} {\bibinfo {author} {\bibfnamefont {A.}~\bibnamefont
  {Derevianko}},\ }\href {https://doi.org/10.1103/PhysRevLett.85.1618}
  {\bibfield  {journal} {\bibinfo  {journal} {Phys. Rev. Lett.}\ }\textbf
  {\bibinfo {volume} {85}},\ \bibinfo {pages} {1618} (\bibinfo {year}
  {2000})}\BibitemShut {NoStop}%
\bibitem [{\citenamefont {Dzuba}\ \emph {et~al.}(2001)\citenamefont {Dzuba},
  \citenamefont {Flambaum},\ and\ \citenamefont {Ginges}}]{DzubaFG01}%
  \BibitemOpen
  \bibfield  {author} {\bibinfo {author} {\bibfnamefont {V.~A.}\ \bibnamefont
  {Dzuba}}, \bibinfo {author} {\bibfnamefont {V.~V.}\ \bibnamefont
  {Flambaum}},\ and\ \bibinfo {author} {\bibfnamefont {J.~S.~M.}\ \bibnamefont
  {Ginges}},\ }\href {https://doi.org/10.1103/PhysRevA.63.062101} {\bibfield
  {journal} {\bibinfo  {journal} {Phys. Rev. A}\ }\textbf {\bibinfo {volume}
  {63}},\ \bibinfo {pages} {062101} (\bibinfo {year} {2001})}\BibitemShut
  {NoStop}%
\bibitem [{\citenamefont {Johnson}\ \emph {et~al.}(2001)\citenamefont
  {Johnson}, \citenamefont {Bednyakov},\ and\ \citenamefont
  {Soff}}]{JohnsonBS01}%
  \BibitemOpen
  \bibfield  {author} {\bibinfo {author} {\bibfnamefont {W.~R.}\ \bibnamefont
  {Johnson}}, \bibinfo {author} {\bibfnamefont {I.}~\bibnamefont {Bednyakov}},\
  and\ \bibinfo {author} {\bibfnamefont {G.}~\bibnamefont {Soff}},\ }\href
  {https://doi.org/10.1103/PhysRevLett.87.233001} {\bibfield  {journal}
  {\bibinfo  {journal} {Phys. Rev. Lett.}\ }\textbf {\bibinfo {volume} {87}},\
  \bibinfo {pages} {233001} (\bibinfo {year} {2001})}\BibitemShut {NoStop}%
\bibitem [{\citenamefont {Kozlov}\ \emph {et~al.}(2001)\citenamefont {Kozlov},
  \citenamefont {Porsev},\ and\ \citenamefont {Tupitsyn}}]{KozlovPT01}%
  \BibitemOpen
  \bibfield  {author} {\bibinfo {author} {\bibfnamefont {M.~G.}\ \bibnamefont
  {Kozlov}}, \bibinfo {author} {\bibfnamefont {S.~G.}\ \bibnamefont {Porsev}},\
  and\ \bibinfo {author} {\bibfnamefont {I.~I.}\ \bibnamefont {Tupitsyn}},\
  }\href {https://doi.org/10.1103/PhysRevLett.86.3260} {\bibfield  {journal}
  {\bibinfo  {journal} {Phys. Rev. Lett.}\ }\textbf {\bibinfo {volume} {86}},\
  \bibinfo {pages} {3260} (\bibinfo {year} {2001})}\BibitemShut {NoStop}%
\bibitem [{\citenamefont {Dzuba}\ \emph {et~al.}(2002)\citenamefont {Dzuba},
  \citenamefont {Flambaum},\ and\ \citenamefont {Ginges}}]{DzubaFG02}%
  \BibitemOpen
  \bibfield  {author} {\bibinfo {author} {\bibfnamefont {V.~A.}\ \bibnamefont
  {Dzuba}}, \bibinfo {author} {\bibfnamefont {V.~V.}\ \bibnamefont
  {Flambaum}},\ and\ \bibinfo {author} {\bibfnamefont {J.~S.~M.}\ \bibnamefont
  {Ginges}},\ }\href {https://doi.org/10.1103/PhysRevD.66.076013} {\bibfield
  {journal} {\bibinfo  {journal} {Phys. Rev. D}\ }\textbf {\bibinfo {volume}
  {66}},\ \bibinfo {pages} {076013} (\bibinfo {year} {2002})}\BibitemShut
  {NoStop}%
\bibitem [{\citenamefont {Flambaum}\ and\ \citenamefont
  {Ginges}(2005)}]{FlambaumG05}%
  \BibitemOpen
  \bibfield  {author} {\bibinfo {author} {\bibfnamefont {V.~V.}\ \bibnamefont
  {Flambaum}}\ and\ \bibinfo {author} {\bibfnamefont {J.~S.~M.}\ \bibnamefont
  {Ginges}},\ }\href {https://doi.org/10.1103/PhysRevA.72.052115} {\bibfield
  {journal} {\bibinfo  {journal} {Phys. Rev. A}\ }\textbf {\bibinfo {volume}
  {72}},\ \bibinfo {pages} {052115} (\bibinfo {year} {2005})}\BibitemShut
  {NoStop}%
\bibitem [{\citenamefont {Porsev}\ \emph {et~al.}(2009)\citenamefont {Porsev},
  \citenamefont {Beloy},\ and\ \citenamefont {Derevianko}}]{PorsevBD09}%
  \BibitemOpen
  \bibfield  {author} {\bibinfo {author} {\bibfnamefont {S.~G.}\ \bibnamefont
  {Porsev}}, \bibinfo {author} {\bibfnamefont {K.}~\bibnamefont {Beloy}},\ and\
  \bibinfo {author} {\bibfnamefont {A.}~\bibnamefont {Derevianko}},\ }\href
  {https://doi.org/10.1103/PhysRevLett.102.181601} {\bibfield  {journal}
  {\bibinfo  {journal} {Phys. Rev. Lett.}\ }\textbf {\bibinfo {volume} {102}},\
  \bibinfo {pages} {181601} (\bibinfo {year} {2009})}\BibitemShut {NoStop}%
\bibitem [{\citenamefont {Porsev}\ \emph {et~al.}(2010)\citenamefont {Porsev},
  \citenamefont {Beloy},\ and\ \citenamefont {Derevianko}}]{PorsevBD10}%
  \BibitemOpen
  \bibfield  {author} {\bibinfo {author} {\bibfnamefont {S.~G.}\ \bibnamefont
  {Porsev}}, \bibinfo {author} {\bibfnamefont {K.}~\bibnamefont {Beloy}},\ and\
  \bibinfo {author} {\bibfnamefont {A.}~\bibnamefont {Derevianko}},\ }\href
  {https://doi.org/10.1103/PhysRevD.82.036008} {\bibfield  {journal} {\bibinfo
  {journal} {Phys. Rev. D}\ }\textbf {\bibinfo {volume} {82}},\ \bibinfo
  {pages} {036008} (\bibinfo {year} {2010})}\BibitemShut {NoStop}%
\bibitem [{\citenamefont {Dzuba}\ \emph {et~al.}(2012)\citenamefont {Dzuba},
  \citenamefont {Berengut}, \citenamefont {Flambaum},\ and\ \citenamefont
  {Roberts}}]{DzubaBFR12}%
  \BibitemOpen
  \bibfield  {author} {\bibinfo {author} {\bibfnamefont {V.~A.}\ \bibnamefont
  {Dzuba}}, \bibinfo {author} {\bibfnamefont {J.~C.}\ \bibnamefont {Berengut}},
  \bibinfo {author} {\bibfnamefont {V.~V.}\ \bibnamefont {Flambaum}},\ and\
  \bibinfo {author} {\bibfnamefont {B.}~\bibnamefont {Roberts}},\ }\href
  {https://doi.org/10.1103/PhysRevLett.109.203003} {\bibfield  {journal}
  {\bibinfo  {journal} {Phys. Rev. Lett.}\ }\textbf {\bibinfo {volume} {109}},\
  \bibinfo {pages} {203003} (\bibinfo {year} {2012})}\BibitemShut {NoStop}%
\bibitem [{\citenamefont {Roberts}\ \emph {et~al.}(2013)\citenamefont
  {Roberts}, \citenamefont {Dzuba},\ and\ \citenamefont
  {Flambaum}}]{RobertsDF2013}%
  \BibitemOpen
  \bibfield  {author} {\bibinfo {author} {\bibfnamefont {B.~M.}\ \bibnamefont
  {Roberts}}, \bibinfo {author} {\bibfnamefont {V.~A.}\ \bibnamefont {Dzuba}},\
  and\ \bibinfo {author} {\bibfnamefont {V.~V.}\ \bibnamefont {Flambaum}},\
  }\href {https://doi.org/10.1103/PhysRevA.87.054502} {\bibfield  {journal}
  {\bibinfo  {journal} {Phys. Rev. A}\ }\textbf {\bibinfo {volume} {87}},\
  \bibinfo {pages} {054502} (\bibinfo {year} {2013})}\BibitemShut {NoStop}%
\bibitem [{\citenamefont {Ginges}\ \emph {et~al.}(2017)\citenamefont {Ginges},
  \citenamefont {Volotka},\ and\ \citenamefont {Fritzsche}}]{GingesVF17}%
  \BibitemOpen
  \bibfield  {author} {\bibinfo {author} {\bibfnamefont {J.~S.~M.}\
  \bibnamefont {Ginges}}, \bibinfo {author} {\bibfnamefont {A.~V.}\
  \bibnamefont {Volotka}},\ and\ \bibinfo {author} {\bibfnamefont
  {S.}~\bibnamefont {Fritzsche}},\ }\href
  {https://doi.org/10.1103/PhysRevA.96.062502} {\bibfield  {journal} {\bibinfo
  {journal} {Phys. Rev. A}\ }\textbf {\bibinfo {volume} {96}},\ \bibinfo
  {pages} {062502} (\bibinfo {year} {2017})}\BibitemShut {NoStop}%
\bibitem [{\citenamefont {Ginges}\ and\ \citenamefont
  {Volotka}(2018)}]{GingesV18}%
  \BibitemOpen
  \bibfield  {author} {\bibinfo {author} {\bibfnamefont {J.~S.~M.}\
  \bibnamefont {Ginges}}\ and\ \bibinfo {author} {\bibfnamefont {A.~V.}\
  \bibnamefont {Volotka}},\ }\href {https://doi.org/10.1103/PhysRevA.98.032504}
  {\bibfield  {journal} {\bibinfo  {journal} {Phys. Rev. A}\ }\textbf {\bibinfo
  {volume} {98}},\ \bibinfo {pages} {032504} (\bibinfo {year}
  {2018})}\BibitemShut {NoStop}%
\bibitem [{\citenamefont {Grunefeld}\ \emph {et~al.}(2019)\citenamefont
  {Grunefeld}, \citenamefont {Roberts},\ and\ \citenamefont
  {Ginges}}]{GrunefeldRG2019}%
  \BibitemOpen
  \bibfield  {author} {\bibinfo {author} {\bibfnamefont {S.~J.}\ \bibnamefont
  {Grunefeld}}, \bibinfo {author} {\bibfnamefont {B.~M.}\ \bibnamefont
  {Roberts}},\ and\ \bibinfo {author} {\bibfnamefont {J.~S.~M.}\ \bibnamefont
  {Ginges}},\ }\href {https://doi.org/10.1103/PhysRevA.100.042506} {\bibfield
  {journal} {\bibinfo  {journal} {Phys. Rev. A}\ }\textbf {\bibinfo {volume}
  {100}},\ \bibinfo {pages} {042506} (\bibinfo {year} {2019})}\BibitemShut
  {NoStop}%
\bibitem [{\citenamefont {Gilbert}\ \emph {et~al.}(1983)\citenamefont
  {Gilbert}, \citenamefont {Watts},\ and\ \citenamefont
  {Wieman}}]{GilbertWW83}%
  \BibitemOpen
  \bibfield  {author} {\bibinfo {author} {\bibfnamefont {S.~L.}\ \bibnamefont
  {Gilbert}}, \bibinfo {author} {\bibfnamefont {R.~N.}\ \bibnamefont {Watts}},\
  and\ \bibinfo {author} {\bibfnamefont {C.~E.}\ \bibnamefont {Wieman}},\
  }\href {https://doi.org/10.1103/PhysRevA.27.581} {\bibfield  {journal}
  {\bibinfo  {journal} {Phys. Rev. A}\ }\textbf {\bibinfo {volume} {27}},\
  \bibinfo {pages} {581} (\bibinfo {year} {1983})}\BibitemShut {NoStop}%
\bibitem [{\citenamefont {Fendel}\ \emph {et~al.}(2007)\citenamefont {Fendel},
  \citenamefont {Bergeson}, \citenamefont {Udem},\ and\ \citenamefont
  {H\"{a}nsch}}]{FendelBUH07}%
  \BibitemOpen
  \bibfield  {author} {\bibinfo {author} {\bibfnamefont {P.}~\bibnamefont
  {Fendel}}, \bibinfo {author} {\bibfnamefont {S.~D.}\ \bibnamefont
  {Bergeson}}, \bibinfo {author} {\bibfnamefont {T.}~\bibnamefont {Udem}},\
  and\ \bibinfo {author} {\bibfnamefont {T.~W.}\ \bibnamefont {H\"{a}nsch}},\
  }\href {https://doi.org/10.1364/OL.32.000701} {\bibfield  {journal} {\bibinfo
   {journal} {Opt. Lett.}\ }\textbf {\bibinfo {volume} {32}},\ \bibinfo {pages}
  {701} (\bibinfo {year} {2007})}\BibitemShut {NoStop}%
\bibitem [{\citenamefont {Stalnaker}\ \emph {et~al.}(2010)\citenamefont
  {Stalnaker}, \citenamefont {Mbele}, \citenamefont {Gerginov}, \citenamefont
  {Fortier}, \citenamefont {Diddams}, \citenamefont {Hollberg},\ and\
  \citenamefont {Tanner}}]{StalnakerMGFDHT10}%
  \BibitemOpen
  \bibfield  {author} {\bibinfo {author} {\bibfnamefont {J.~E.}\ \bibnamefont
  {Stalnaker}}, \bibinfo {author} {\bibfnamefont {V.}~\bibnamefont {Mbele}},
  \bibinfo {author} {\bibfnamefont {V.}~\bibnamefont {Gerginov}}, \bibinfo
  {author} {\bibfnamefont {T.~M.}\ \bibnamefont {Fortier}}, \bibinfo {author}
  {\bibfnamefont {S.~A.}\ \bibnamefont {Diddams}}, \bibinfo {author}
  {\bibfnamefont {L.}~\bibnamefont {Hollberg}},\ and\ \bibinfo {author}
  {\bibfnamefont {C.~E.}\ \bibnamefont {Tanner}},\ }\href
  {https://doi.org/10.1103/PhysRevA.81.043840} {\bibfield  {journal} {\bibinfo
  {journal} {Phys. Rev. A}\ }\textbf {\bibinfo {volume} {81}},\ \bibinfo
  {pages} {043840} (\bibinfo {year} {2010})}\BibitemShut {NoStop}%
\bibitem [{\citenamefont {Wu}\ \emph {et~al.}(2013)\citenamefont {Wu},
  \citenamefont {Liu}, \citenamefont {Wu}, \citenamefont {Lee},\ and\
  \citenamefont {Cheng}}]{WuLWLC13}%
  \BibitemOpen
  \bibfield  {author} {\bibinfo {author} {\bibfnamefont {C.-M.}\ \bibnamefont
  {Wu}}, \bibinfo {author} {\bibfnamefont {T.-W.}\ \bibnamefont {Liu}},
  \bibinfo {author} {\bibfnamefont {M.-H.}\ \bibnamefont {Wu}}, \bibinfo
  {author} {\bibfnamefont {R.-K.}\ \bibnamefont {Lee}},\ and\ \bibinfo {author}
  {\bibfnamefont {W.-Y.}\ \bibnamefont {Cheng}},\ }\href
  {https://doi.org/10.1364/OL.38.003186} {\bibfield  {journal} {\bibinfo
  {journal} {Opt. Lett.}\ }\textbf {\bibinfo {volume} {38}},\ \bibinfo {pages}
  {3186} (\bibinfo {year} {2013})}\BibitemShut {NoStop}%
\bibitem [{\citenamefont {Jin}\ \emph {et~al.}(2013)\citenamefont {Jin},
  \citenamefont {Zhang}, \citenamefont {Xiang}, \citenamefont {Wang},
  \citenamefont {Ma}, \citenamefont {Zhao}, \citenamefont {Xiao},\ and\
  \citenamefont {Jia}}]{JinZXWMZXJ13}%
  \BibitemOpen
  \bibfield  {author} {\bibinfo {author} {\bibfnamefont {L.}~\bibnamefont
  {Jin}}, \bibinfo {author} {\bibfnamefont {Y.-C.}\ \bibnamefont {Zhang}},
  \bibinfo {author} {\bibfnamefont {S.-S.}\ \bibnamefont {Xiang}}, \bibinfo
  {author} {\bibfnamefont {L.-R.}\ \bibnamefont {Wang}}, \bibinfo {author}
  {\bibfnamefont {J.}~\bibnamefont {Ma}}, \bibinfo {author} {\bibfnamefont
  {Y.-T.}\ \bibnamefont {Zhao}}, \bibinfo {author} {\bibfnamefont {L.-T.}\
  \bibnamefont {Xiao}},\ and\ \bibinfo {author} {\bibfnamefont {S.-T.}\
  \bibnamefont {Jia}},\ }\href
  {http://stacks.iop.org/0256-307X/30/i=10/a=103201} {\bibfield  {journal}
  {\bibinfo  {journal} {Chinese Physics Letters}\ }\textbf {\bibinfo {volume}
  {30}},\ \bibinfo {pages} {103201} (\bibinfo {year} {2013})}\BibitemShut
  {NoStop}%
\bibitem [{\citenamefont {Yang}\ \emph {et~al.}(2016)\citenamefont {Yang},
  \citenamefont {Wang}, \citenamefont {Yang},\ and\ \citenamefont
  {Wang}}]{YangWYW16}%
  \BibitemOpen
  \bibfield  {author} {\bibinfo {author} {\bibfnamefont {G.}~\bibnamefont
  {Yang}}, \bibinfo {author} {\bibfnamefont {J.}~\bibnamefont {Wang}}, \bibinfo
  {author} {\bibfnamefont {B.}~\bibnamefont {Yang}},\ and\ \bibinfo {author}
  {\bibfnamefont {J.}~\bibnamefont {Wang}},\ }\href
  {http://stacks.iop.org/1612-202X/13/i=8/a=085702} {\bibfield  {journal}
  {\bibinfo  {journal} {Laser Physics Letters}\ }\textbf {\bibinfo {volume}
  {13}},\ \bibinfo {pages} {085702} (\bibinfo {year} {2016})}\BibitemShut
  {NoStop}%
\bibitem [{\citenamefont {Tsekeris}\ \emph {et~al.}(1974)\citenamefont
  {Tsekeris}, \citenamefont {Gupta}, \citenamefont {Happer}, \citenamefont
  {Belin},\ and\ \citenamefont {Svanberg}}]{TsekerisGHBS1974}%
  \BibitemOpen
  \bibfield  {author} {\bibinfo {author} {\bibfnamefont {P.}~\bibnamefont
  {Tsekeris}}, \bibinfo {author} {\bibfnamefont {R.}~\bibnamefont {Gupta}},
  \bibinfo {author} {\bibfnamefont {W.}~\bibnamefont {Happer}}, \bibinfo
  {author} {\bibfnamefont {G.}~\bibnamefont {Belin}},\ and\ \bibinfo {author}
  {\bibfnamefont {S.}~\bibnamefont {Svanberg}},\ }\href
  {https://doi.org/https://doi.org/10.1016/0375-9601(74)90418-6} {\bibfield
  {journal} {\bibinfo  {journal} {Physics Letters A}\ }\textbf {\bibinfo
  {volume} {48}},\ \bibinfo {pages} {101 } (\bibinfo {year}
  {1974})}\BibitemShut {NoStop}%
\bibitem [{\citenamefont {Tsekeris}\ and\ \citenamefont
  {Gupta}(1975)}]{TsekerisG1975}%
  \BibitemOpen
  \bibfield  {author} {\bibinfo {author} {\bibfnamefont {P.}~\bibnamefont
  {Tsekeris}}\ and\ \bibinfo {author} {\bibfnamefont {R.}~\bibnamefont
  {Gupta}},\ }\href {https://doi.org/10.1103/PhysRevA.11.455} {\bibfield
  {journal} {\bibinfo  {journal} {Phys. Rev. A}\ }\textbf {\bibinfo {volume}
  {11}},\ \bibinfo {pages} {455} (\bibinfo {year} {1975})}\BibitemShut
  {NoStop}%
\bibitem [{\citenamefont {Tsekeris}(1976)}]{Tsekeris1976}%
  \BibitemOpen
  \bibfield  {author} {\bibinfo {author} {\bibfnamefont {P.}~\bibnamefont
  {Tsekeris}},\ }\emph {\bibinfo {title} {Hyperfine-Structure Measurements in
  Excited States of Alkali Atoms by cw Laser Spectroscopy}},\ \href@noop {}
  {Ph.D. thesis},\ \bibinfo  {school} {Columbia University} (\bibinfo {year}
  {1976})\BibitemShut {NoStop}%
\bibitem [{\citenamefont {Farley}\ \emph {et~al.}(1977)\citenamefont {Farley},
  \citenamefont {Tsekeris},\ and\ \citenamefont {Gupta}}]{FarleyTG1977}%
  \BibitemOpen
  \bibfield  {author} {\bibinfo {author} {\bibfnamefont {J.}~\bibnamefont
  {Farley}}, \bibinfo {author} {\bibfnamefont {P.}~\bibnamefont {Tsekeris}},\
  and\ \bibinfo {author} {\bibfnamefont {R.}~\bibnamefont {Gupta}},\ }\href
  {https://doi.org/10.1103/PhysRevA.15.1530} {\bibfield  {journal} {\bibinfo
  {journal} {Phys. Rev. A}\ }\textbf {\bibinfo {volume} {15}},\ \bibinfo
  {pages} {1530} (\bibinfo {year} {1977})}\BibitemShut {NoStop}%
\bibitem [{\citenamefont {Antypas}\ and\ \citenamefont
  {Elliott}(2013)}]{antypasm12013}%
  \BibitemOpen
  \bibfield  {author} {\bibinfo {author} {\bibfnamefont {D.}~\bibnamefont
  {Antypas}}\ and\ \bibinfo {author} {\bibfnamefont {D.~S.}\ \bibnamefont
  {Elliott}},\ }\href {https://doi.org/10.1103/PhysRevA.87.042505} {\bibfield
  {journal} {\bibinfo  {journal} {Phys. Rev. A}\ }\textbf {\bibinfo {volume}
  {87}},\ \bibinfo {pages} {042505} (\bibinfo {year} {2013})}\BibitemShut
  {NoStop}%
\bibitem [{\citenamefont {Choi}\ and\ \citenamefont
  {Elliott}(2016)}]{choipnc2016}%
  \BibitemOpen
  \bibfield  {author} {\bibinfo {author} {\bibfnamefont {J.}~\bibnamefont
  {Choi}}\ and\ \bibinfo {author} {\bibfnamefont {D.~S.}\ \bibnamefont
  {Elliott}},\ }\href {https://doi.org/10.1103/PhysRevA.93.023432} {\bibfield
  {journal} {\bibinfo  {journal} {Phys. Rev. A}\ }\textbf {\bibinfo {volume}
  {93}},\ \bibinfo {pages} {023432} (\bibinfo {year} {2016})}\BibitemShut
  {NoStop}%
\bibitem [{\citenamefont {Toh}\ \emph {et~al.}(2019)\citenamefont {Toh},
  \citenamefont {Damitz}, \citenamefont {Tanner}, \citenamefont {Johnson},\
  and\ \citenamefont {Elliott}}]{TohDTJE19}%
  \BibitemOpen
  \bibfield  {author} {\bibinfo {author} {\bibfnamefont {G.}~\bibnamefont
  {Toh}}, \bibinfo {author} {\bibfnamefont {A.}~\bibnamefont {Damitz}},
  \bibinfo {author} {\bibfnamefont {C.~E.}\ \bibnamefont {Tanner}}, \bibinfo
  {author} {\bibfnamefont {W.~R.}\ \bibnamefont {Johnson}},\ and\ \bibinfo
  {author} {\bibfnamefont {D.~S.}\ \bibnamefont {Elliott}},\ }\href
  {https://doi.org/10.1103/PhysRevLett.123.073002} {\bibfield  {journal}
  {\bibinfo  {journal} {Phys. Rev. Lett.}\ }\textbf {\bibinfo {volume} {123}},\
  \bibinfo {pages} {073002} (\bibinfo {year} {2019})}\BibitemShut {NoStop}%
\bibitem [{\citenamefont {Barakhshan}\ \emph {et~al.}(2021)\citenamefont
  {Barakhshan}, \citenamefont {Marrs}, \citenamefont {Arora}, \citenamefont
  {Eigenmann},\ and\ \citenamefont {Safronova}}]{safronovadatabase}%
  \BibitemOpen
  \bibfield  {author} {\bibinfo {author} {\bibfnamefont {P.}~\bibnamefont
  {Barakhshan}}, \bibinfo {author} {\bibfnamefont {A.}~\bibnamefont {Marrs}},
  \bibinfo {author} {\bibfnamefont {B.}~\bibnamefont {Arora}}, \bibinfo
  {author} {\bibfnamefont {R.}~\bibnamefont {Eigenmann}},\ and\ \bibinfo
  {author} {\bibfnamefont {M.~S.}\ \bibnamefont {Safronova}},\ }\href@noop {}
  {}\bibinfo {howpublished} {{\textit{Portal for High-Precision Atomic Data and
  Computation}} (version 1.0). University of Delaware, Newark, DE, USA. URL:
  {\tt{https://www.udel.edu/atom}} [January 2021].} (\bibinfo {year}
  {2021})\BibitemShut {NoStop}%
\bibitem [{\citenamefont {Anderson}\ \emph {et~al.}(1990)\citenamefont
  {Anderson}, \citenamefont {Jones}, \citenamefont {Cooper}, \citenamefont
  {Smith}, \citenamefont {Elliott}, \citenamefont {Ritsch},\ and\ \citenamefont
  {Zoller}}]{AndersonJCSERZ1990}%
  \BibitemOpen
  \bibfield  {author} {\bibinfo {author} {\bibfnamefont {M.~H.}\ \bibnamefont
  {Anderson}}, \bibinfo {author} {\bibfnamefont {R.~D.}\ \bibnamefont {Jones}},
  \bibinfo {author} {\bibfnamefont {J.}~\bibnamefont {Cooper}}, \bibinfo
  {author} {\bibfnamefont {S.~J.}\ \bibnamefont {Smith}}, \bibinfo {author}
  {\bibfnamefont {D.~S.}\ \bibnamefont {Elliott}}, \bibinfo {author}
  {\bibfnamefont {H.}~\bibnamefont {Ritsch}},\ and\ \bibinfo {author}
  {\bibfnamefont {P.}~\bibnamefont {Zoller}},\ }\href
  {https://doi.org/10.1103/PhysRevA.42.6690} {\bibfield  {journal} {\bibinfo
  {journal} {Phys. Rev. A}\ }\textbf {\bibinfo {volume} {42}},\ \bibinfo
  {pages} {6690} (\bibinfo {year} {1990})}\BibitemShut {NoStop}%
\bibitem [{\citenamefont {Neil}\ and\ \citenamefont
  {Atkinson}(1984)}]{NeilA84}%
  \BibitemOpen
  \bibfield  {author} {\bibinfo {author} {\bibfnamefont {W.~S.}\ \bibnamefont
  {Neil}}\ and\ \bibinfo {author} {\bibfnamefont {J.~B.}\ \bibnamefont
  {Atkinson}},\ }\href@noop {} {\bibfield  {journal} {\bibinfo  {journal}
  {Journal of Physics B: Atomic and Molecular Physics}\ }\textbf {\bibinfo
  {volume} {17}},\ \bibinfo {pages} {693} (\bibinfo {year} {1984})}\BibitemShut
  {NoStop}%
\bibitem [{\citenamefont {Weber}\ and\ \citenamefont
  {Sansonetti}(1987)}]{WeberS1987}%
  \BibitemOpen
  \bibfield  {author} {\bibinfo {author} {\bibfnamefont {K.-H.}\ \bibnamefont
  {Weber}}\ and\ \bibinfo {author} {\bibfnamefont {C.~J.}\ \bibnamefont
  {Sansonetti}},\ }\href {https://doi.org/10.1103/PhysRevA.35.4650} {\bibfield
  {journal} {\bibinfo  {journal} {Phys. Rev. A}\ }\textbf {\bibinfo {volume}
  {35}},\ \bibinfo {pages} {4650} (\bibinfo {year} {1987})}\BibitemShut
  {NoStop}%
\bibitem [{\citenamefont {Tang}\ \emph {et~al.}(2019)\citenamefont {Tang},
  \citenamefont {Lou},\ and\ \citenamefont {Shi}}]{TangLS2019}%
  \BibitemOpen
  \bibfield  {author} {\bibinfo {author} {\bibfnamefont {Y.-B.}\ \bibnamefont
  {Tang}}, \bibinfo {author} {\bibfnamefont {B.-Q.}\ \bibnamefont {Lou}},\ and\
  \bibinfo {author} {\bibfnamefont {T.-Y.}\ \bibnamefont {Shi}},\ }\href
  {https://doi.org/10.1088/1361-6455/ac1329} {\bibfield  {journal} {\bibinfo
  {journal} {Journal of Physics B: Atomic, Molecular and Optical Physics}\
  }\textbf {\bibinfo {volume} {52}},\ \bibinfo {pages} {055002} (\bibinfo
  {year} {2019})}\BibitemShut {NoStop}%
\bibitem [{\citenamefont {Svanberg}\ and\ \citenamefont
  {Belin}(1974)}]{SvanbergB1974}%
  \BibitemOpen
  \bibfield  {author} {\bibinfo {author} {\bibfnamefont {S.}~\bibnamefont
  {Svanberg}}\ and\ \bibinfo {author} {\bibfnamefont {G.}~\bibnamefont
  {Belin}},\ }\href {https://doi.org/10.1088/0022-3700/7/3/024} {\bibfield
  {journal} {\bibinfo  {journal} {Journal of Physics B: Atomic and Molecular
  Physics}\ }\textbf {\bibinfo {volume} {7}},\ \bibinfo {pages} {L82} (\bibinfo
  {year} {1974})}\BibitemShut {NoStop}%
\bibitem [{\citenamefont {Deech}\ \emph {et~al.}(1977)\citenamefont {Deech},
  \citenamefont {Luypaert}, \citenamefont {Pendrill},\ and\ \citenamefont
  {Series}}]{DeechLPS1977}%
  \BibitemOpen
  \bibfield  {author} {\bibinfo {author} {\bibfnamefont {J.}~\bibnamefont
  {Deech}}, \bibinfo {author} {\bibfnamefont {R.}~\bibnamefont {Luypaert}},
  \bibinfo {author} {\bibfnamefont {L.}~\bibnamefont {Pendrill}},\ and\
  \bibinfo {author} {\bibfnamefont {G.}~\bibnamefont {Series}},\ }\href
  {https://doi.org/10.1088/0022-3700/10/5/004} {\bibfield  {journal} {\bibinfo
  {journal} {Journal of Physics B: Atomic and Molecular Physics}\ }\textbf
  {\bibinfo {volume} {10}},\ \bibinfo {pages} {L137} (\bibinfo {year}
  {1977})}\BibitemShut {NoStop}%
\end{thebibliography}%

\end{document}